\documentclass[reprint, twocolumn, aps]{revtex4-2}
\usepackage{hyperref}
\usepackage{color}
\usepackage{multirow}
\usepackage{epsfig}
\usepackage{graphicx}
\usepackage{amsmath}
\usepackage{amssymb}
\usepackage{bm}
\usepackage{dcolumn}
\usepackage{braket}
\usepackage{bbold}
\usepackage{ulem}

\usepackage{graphicx}
\usepackage{dcolumn}
\usepackage{bm}
\usepackage{diagbox}
\usepackage{epsfig}
\usepackage{tabularx}
\usepackage{url}
\usepackage{chngcntr} 

\begin{document}

\preprint{APS/123-QED}

\title{Bayesian optimization with active learning of Ta-Nb-Hf-Zr-Ti system for spin transport properties}

\author{Ruiwen Xie}
\email{ruiwen.xie@tmm.tu-darmstadt.de}
\affiliation{Institute of Materials Science, Technical University of Darmstadt, Darmstadt, Germany}
\author{Yixuan Zhang}
\affiliation{Institute of Materials Science, Technical University of Darmstadt, Darmstadt, Germany}

\author{Fu Li}
\affiliation{Institute of Materials Science, Technical University of Darmstadt, Darmstadt, Germany}

\author{Zhiyuan Li}
\affiliation{Institute of Materials Science, Technical University of Darmstadt, Darmstadt, Germany}
 
\author{Hongbin Zhang}%
\affiliation{Institute of Materials Science, Technical University of Darmstadt, Darmstadt, Germany}%


\date{\today}

\begin{abstract}

Designing materials with enhanced spin charge conversion, i.e., with high spin Hall conductivity (SHC) and low longitudinal electric conductivity (hence large spin Hall angle (SHA)), is a challenging task, especially in the presence of a vast chemical space for compositionally complex alloys (CCAs). In this work, focusing on the Ta-Nb-Hf-Zr-Ti system, we confirm that CCAs exhibit significant spin Hall conductivities and propose a multi-objective Bayesian optimization approach (MOBO) incorporated with active learning (AL) in order to screen for the optimal compositions with significant SHC and SHA. As a result, within less than 5 iterations we are able to target the TaZr-dominated systems displaying both high magnitudes of SHC ($\sim$-2.0 $(10^{-3}~\Omega~\mathrm{cm})^{-1}$) and SHA ($\sim$0.03). The SHC is mainly ascribed to the extrinsic skew scattering mechanism. Our work provides an efficient route for identifying new materials with significant SHE, which can be straightforwardly generalized to optimize other properties in a vast chemical space.

\end{abstract}

\maketitle



\section{\label{sec:intro}Introduction}

The spin Hall effect (SHE), which describes a phenomenon that a transverse spin imbalance (spin Hall voltage) is induced by a longitudinal charge current, is an intriguing subject in the development of spintronic devices.~\cite{hirsch1999spin} The spin current generation can be realized in nonmagnetic materials, in distinction with the injection of spin polarized current using ferromagnetic materials. In addition, for a high efficiency of charge to spin conversion, a large spin Hall angle (SHA) is required. Large SHAs have been predicted theoretically and found experimentally in a number of materials. For instance, the measured SHA of Au wires is 0.1~\cite{seki2008giant} and 0.08 for Pt.~\cite{ando2008electric, liu2011spin} Even higher SHAs have also been found in $\beta$-Ta (-0.15)~\cite{liu2012spin} and $\beta$-W thin films (0.3).~\cite{pai2012spin} For these single-element systems, the large SHAs can be attributed to the large spin Hall conductivites (SHCs), in particular the intrinsic contribution originated from the atomic spin-orbit coupling (SOC), thus hard to manipulate. The charge to spin conversion has found to be thickness-dependent in perovskite SrIrO$_3$ thin films.~\cite{everhardt2019tunable} More interestingly, the anticorrelation between carrier concentration and resistivity does not hold in SrIrO$_3$ thin films like in topological insulators, producing a large SHA of 0.3-0.5 when the thickness of SrIrO$_3$ films is larger than 10 nm. Besides, the magnitude of SHA as high as -0.59 has been detected in a 5.3-nm-thick $\beta$-Ta$_x$W$_{1-x}$.~\cite{qian2020spin}
High-throughput \textit{ab initio} calculations have also been performed on the intrinsic SHC of over 20,000 nonmagnetic crystals and have identified 11 materials with an SHC comparable to or even larger than that of Pt.~\cite{zhang2021different} However, material design towards manipulating extrinsic contribution to SHE by multi-principal element alloying as in compositionally complex alloys (CCAs), is still lacking. ~\cite{chadova2015tailoring} 

The CCAs, in particular high entropy alloys (HEAs), have attracted much attention in the material science community. The CCAs typically consist of five or more compositions but they are not necessarily HEAs. 
Using the composition-based definition, HEAs are those with “principal elements with the concentration of each element being between 35 and 5 at.\%,” thus possessing high mixing entropy. 
Despite the widely studied mechanical and magnetic properties of HEAs, their transport properties, especially the spin-transport properties, have been scarcely reported. One pioneering work regarding the spin-transport properties of HEA was done by Chen \textit{et al.} on a sputtered Ta$_{24.9}$Nb$_{18.7}$Hf$_{17.7}$Zr$_{18.3}$Ti$_{20.4}$ film.~\cite{chen2017spin} The spin Hall conductivity (SHC) of this HEA was estimated to be $2.9\times10^4 [\hbar/2e]~\Omega^{-1}~m^{-1}$, which is comparable to that of $\beta$-Ta ($6.4\times10^4 [\hbar/2e]~\Omega^{-1}~m^{-1}$).~\cite{liu2012spin} Moreover, its spin-orbit torque efficiency is quite sizeable. As a matter of fact, the refractory HEAs consisting of Ta, Nb, Hf, Zr and Ti display significantly intriguing physical properties. In 2014, the body-centred cubic (bcc) Ta$_{34}$Nb$_{33}$Hf$_8$Zr$_{14}$Ti$_{11}$ was proved to be a phonon-mediated superconductor in the weak electron-phonon coupling limit, implying the existence of a Fermi surface (FS).~\cite{kovzelj2014discovery} Later, it was revealed by high-resolution Compton scattering experiments that FS can survive the extreme chemical disorder in an equiatomic alloy NiFeCoCr.~\cite{robarts2020extreme} 
Moreover, the residual resistivity of TaNbHfZrTi HEAs are rather high, being around 80 $\mu\Omega\cdot$cm, ~\cite{dong2023transport} which is promising for obtaining large SHA. 

Due to the lack of study on the spin transport properties of TaNbHfZrTi CCAs and their prominent potential applications, in this work we are aiming at searching within the Ta-Nb-Hf-Zr-Ti compositional space for materials with simultaneously high SHC and SHA. 
Unfortunately, the five-component Ta-Nb-Hf-Zr-Ti CCAs pose a vast compositional space (approximately 4.6 million combinations with 1 at.\% interval). It would be unaffordable, even for theoretical calculations, to study the whole chemical space. Additionally, the design objectives are multiple in most cases. This requires a more efficient sampling method in the compositional space, such as by guiding the materials search using a surrogate model. Bayesian optimization (BO) of a Gaussian process (GP) is a feasible choice.~\cite{torres2019low,zhang2023autonomous,pedersen2021bayesian} Furthermore, the multi-objective Bayesian optimization (MOBO) methods have been popular in material design since they are able to work well with minimal data. In terms of the optimal design, the MOBO methods employ a heuristic-based search to increase the system's state of knowledge. The MOBO schemes have been used to design Mo-Nb-Ti-V-W systems for application to gas turbine engines with high ductility, high-temperature yield strength, low density, high thermal conductivity, narrow solidification range, high solidus temperature, and a small linear thermal expansion coefficient.~\cite{khatamsaz2023bayesian} In addition, the active-learning (AL) method based on Bayesian analysis can be incorporated into the BO process to reduce the required sampling data.~\cite{rasmussen2006gaussian} And one can stop the AL once the desired properties are achieved.~\cite{li2006confidence} 

In this work we have established a workflow by combining the MOBO and AL, in order to explore the vast chemical space of the Ta-Nb-Hf-Zr-Ti system for compositions with optimal SHE. We have confirmed that this scheme is capable of efficiently revealing the promising candidates in CCAs. 
 
\section{\label{sec:method}Methodology}

\begin{figure}[h!]
	\centering
	\includegraphics[width=1.0\columnwidth]{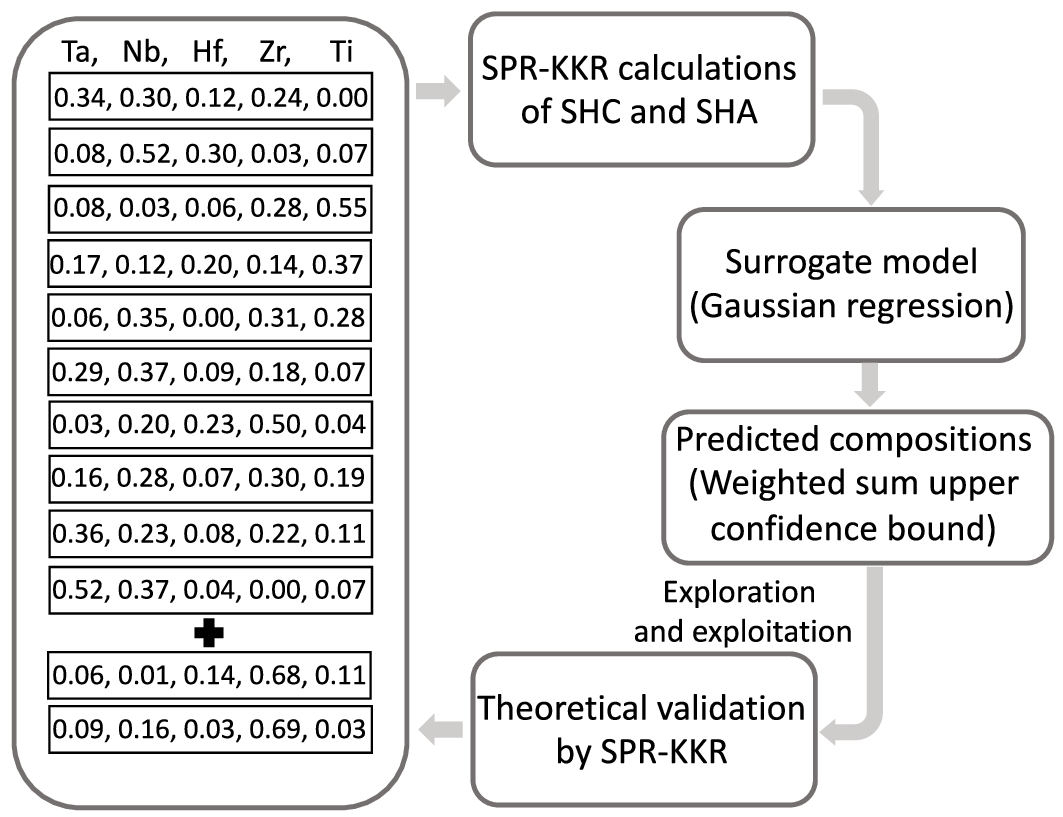}
	\caption{\label{fig:workflow}  Illustration of the workflow of Bayesian optimization combined with active learning for simultaneously high SHC and SHA in Ta-Nb-Hf-Zr-Ti alloys.}
\end{figure}

The BO framework combined with AL was constructed using BoTorch.~\cite{balandat2020botorch} We first generated a complete composition space consisting of Ta, Nb, Hf, Zr and Ti with an concentration interval of 1\%, and for each element its concentration ranges from 0\% to 100\%. In total there are 4598126 different compositions, from which 10 compositional combinations were randomly selected, as displayed in Fig.~\ref{fig:workflow}. Subsequently using the spin-polarized relativistic Korringa-Kohn-Rostoker (SPR-KKR) package,~\cite{ebert2011calculating} the SHC ($\sigma_{xy}^{z}$) and longitudinal electric conductivity ($\sigma_{xx}$) were calculated based on the linear response Kubo-Bastin (KB) formalism.~\cite{kodderitzsch2015linear} Chemical disorder and vertex corrections were treated by the means of coherent potential approximation (CPA).~\cite{butler1985theory,velicky1969theory} In specific, as described in Ref.~\cite{kodderitzsch2015linear}, with the KB formalism implemented in the SPR-KKR package, a fully relativistic Dirac four-component scheme was used for the basis functions. In this work, an angular momentum cutoff of $l_{max} = 4$ was adopted throughout since we treat the 4$f$ electrons of Hf as valence states. The self-consistent field (SCF) potentials were obtained by employing the Vosko-Wilk-Nussair (VWN) parametrization~\cite{vosko1980accurate} for the local density approximation (LDA) exchange-correlation functional. The involved energy integration was performed on a semicircle in the complex plane using 50 energy points and $28^3$ $k$-points in the Brillouin zone (BZ). Using the optimized potentials, subsequent KB transport calculations were performed. For concentrated alloys, $10^6$ $k$-points were sampled in the BZ and the calculations were carried out on the real-energy axis. In contrast, for pure elements, a small imaginary part needs to be added, $z=E_F + i\eta$. Here, we took $\eta=10^{-4}$ and the number of $k$-points in the BZ being $10^6$. With the calculated $\sigma_{xy}^{z}$ and $\sigma_{xx}$, the SHA can be obtained by $\sigma_{xy}^{z}$/$\sigma_{xx}$. We then fitted the Gaussian process regression (GPR) model with the chemical compositions as inputs $X_n=(x_1,...,x_i,...,x_n)^T \in \mathbb{R}^n$ with $ x_i\in[0,1]^d $, where $d = 5$ (number of considered elements). The corresponding outputs are $Y_n=(y_1,...,y_i,...,y_n)^T$ with $y_i = [\mathrm{SHC}, \mathrm{SHA}]$. A FixedNoiseGP and the SpectralDeltaKernel implemented in Gpytorch~\cite{gardner2018gpytorch} were used in the BO algorithm. For the acquisition function, we employed the upper confidence bound (UCB). The UCB balances exploration and exploitation by assigning a score of $\mu+\sqrt{\beta} \cdot \sigma$ ($\mu$ is mean and $\sigma^2$ is variance of the posterior), in which we took $\beta=0.01$ (dubbed as exploitation) and $\beta=100$ (dubbed as exploration) for different purposes. Since in this work we are targeting two objectives, i.e., maximum SHC and maximum SHA, we adopted the weighted sum of UCB, namely by summing up the outcomes of the two UCB functions of SHC and SHA by weights. The weights corresponding to SHC and SHA were determined by the normalized best SHC and SHA values in the dataset so that SHC$_{\mathrm{best}} * w_{\mathrm{SHC}}$ is close to  SHA$_{\mathrm{best}} * w_{\mathrm{SHA}}$. Excluding those compositional combinations already present in the training dataset, the weighted sum UCB was evaluated for all the other compositions, from within, five compositions from exploitation and five from exploration categories were selected. The SHC and SHA of those selected alloys were then calculated using SPR-KKR package and added to the training dataset for initialization of another iteration. The whole process was iterated until the desired properties were identified. 

In this work, all the calculations of Ta-Nb-Hf-Zr-Ti alloys were carried out in body-centred cubic (bcc) phase and the lattice parameters were estimated from the atomic volume of each element observed in experiments by calculating the average with atomic concentration. This averaging method has been validated to be comparable with experimental volume.~\cite{fukushima2022automatic}

\section{\label{sec:result}Results and Discussion}

\begin{figure*}[ht!]
	\centering
	\includegraphics[width=1.0\linewidth]{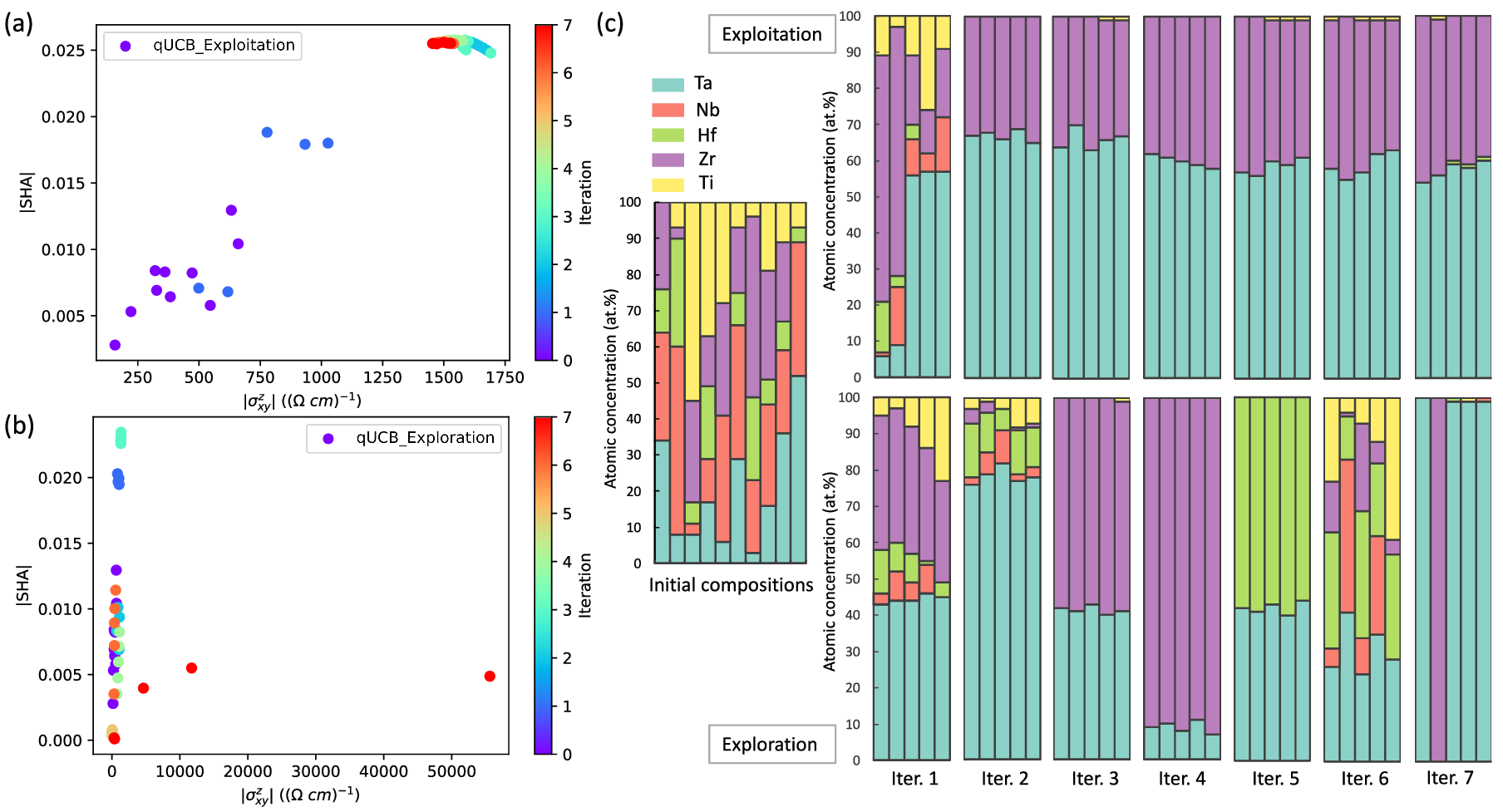}
	\caption{\label{fig:UCB_iter}  Evolutions of $\mid$SHC$\mid$ and $\mid$SHA$\mid$ using UCB acquisition function with (a) $\beta=0.01$ and (b) $\beta=100$. (c) represents the compositional distributions with iteration corresponding to the algorithms adopted in (a) and (b) for exploitation and exploration, respectively. The initial dataset consisting of the 10 random compositions are also shown for reference.}
\end{figure*}

\begin{figure*}[ht!]
	\centering
	\includegraphics[width=1.0\linewidth]{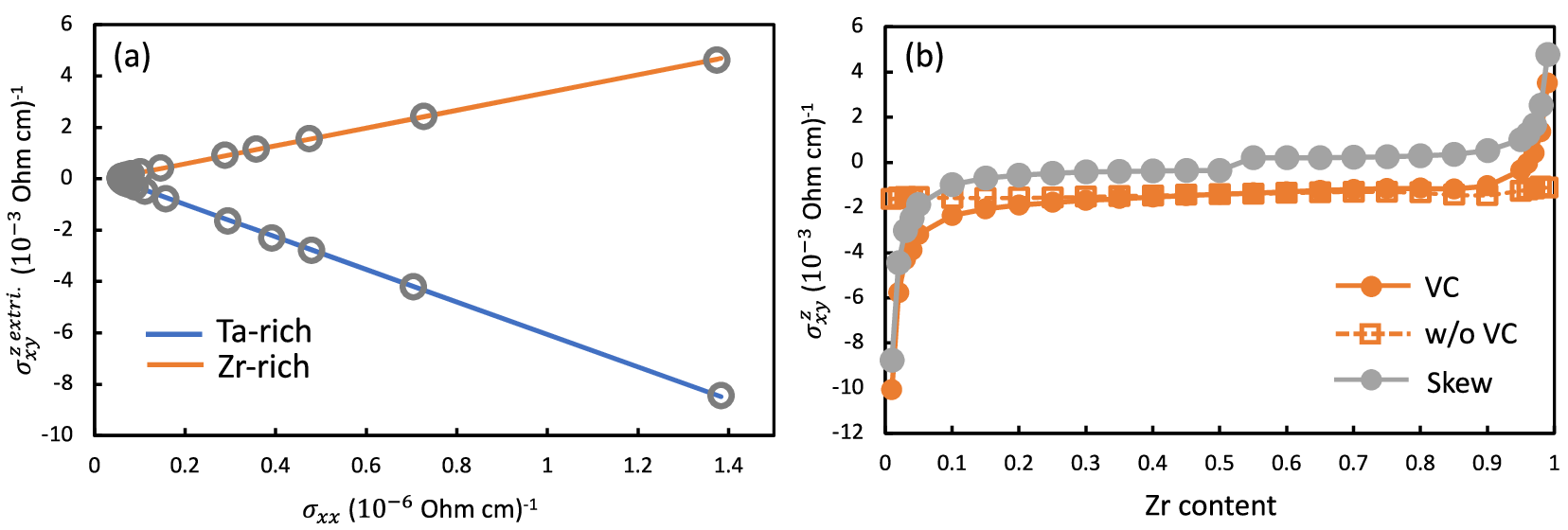}
	\caption{\label{fig:TaZr}  (a) The extrinsic spin Hall conductivity $\sigma_{xy}^{z~extri}$ versus the longitudinal conductivity $\sigma_{xx}$ for Ta$_{1-x}$Zr$_x$. (b) Spin Hall conductivities $\sigma_{xy}^{z}$ for Ta$_{1-x}$Zr$_x$. The orange solid circles and hollow squares represent $\sigma_{xy}^{z}$ with and without vertex corrections,  respectively. Gray solid circles are the contributions from skewness scattering.}
\end{figure*}

\begin{figure*}[ht!]
	\centering
	\includegraphics[width=1.0\linewidth]{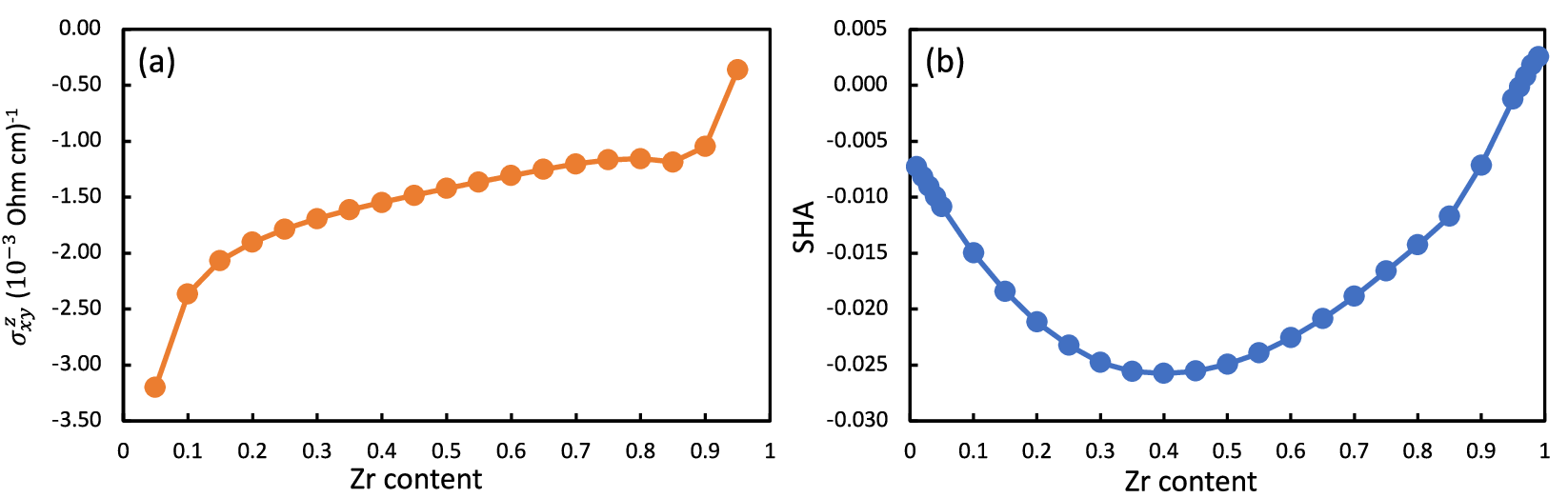}
	\caption{\label{fig:SHC_SHA}  (a) $\sigma_{xy}^{z}$ and (b) SHA ($\sigma_{xy}^{z}/\sigma_{xx}$) as a function of Zr content in Ta$_{1-x}$Zr$_x$. For $\sigma_{xy}^{z}$, we only show $0.05 \leqslant x \leqslant 0.95 $ to make the trend more clear.}
\end{figure*}

Following the workflow demonstrated in Sec.~\ref{sec:method}, the evolutions of $\mid$SHC$\mid$ and $\mid$SHA$\mid$ with number of iterations, as well as the compositional changes from the first to the seventh iteration, are illustrated in Fig.~\ref{fig:UCB_iter}. The vast compositional space motivates us to adopt two regimes for the UCB acquisition function, that is, $\beta=0.01$ for exploitation purpose to approach promising candidates with relatively large mean value, while $\beta=100$ to explore more thoroughly the whole space by selecting chemical compositions showing relatively large variance. It can be seen from Fig.~\ref{fig:UCB_iter} (a) that in the exploitation regime, from the third iteration both $\mid$SHC$\mid$ and $\mid$SHA$\mid$ reach around the maximum of approximately 1700 $(\Omega~ cm)^{-1}$ and 0.025. On the other hand, in the exploration regime, the distributions of $\mid$SHC$\mid$ and $\mid$SHA$\mid$ are rather dispersed (see Fig.~\ref{fig:UCB_iter} (b)). 
The chemical distributions of Ta, Nb, Hf, Zr and Ti from the first to the seventh iteration in the exploitation and the exploration regimes are displayed in Fig.~\ref{fig:UCB_iter} (c). The initial dataset used for training is also displayed. It can be clearly observed that from the second iteration in the exploitation regime Ta and Zr components are dominant, which as a consequence motivates us to investigate more thoroughly the transport properties of Ta-Zr binary alloys. In addition, as can be noticed from Fig.~\ref{fig:UCB_iter} (b), in the seventh iteration of the exploration regime, two compositions display exceptionally high $\mid$SHC$\mid$, which correspond to Ta$_{0.99}$Nb$_{0.01}$ ($\sim$55605 $(\Omega~cm)^{-1}$) and Ta$_{0.99}$Hf $_{0.01}$ ($\sim$11722 $(\Omega~cm)^{-1}$). As a matter of fact, it has been reported that in the dilute limit, SHC depends sensitively on concentration.~\cite{lowitzer2011extrinsic} Furthermore, the SHC has been found to result from a subtle interplay between spin-orbit and potential scattering in different angular-momentum channels.~\cite{herschbach2013insight} Usually large SHC is obtained when the relative difference in the spin-orbit coupling (SOC) strength of the host and impurity increases.~\cite{chadova2015tailoring} 
 
We thus focus on Ta-Zr binary alloys (Ta$_{1-x}$Zr$_x$) and study the concentration dependencies of SHC and SHA.  according to the relation given in Ref.~\cite{lowitzer2011extrinsic}, the solid state materials may be classified according to the scaling relation between the SHC $\sigma_{xy}^{z}$ and the longitudinal conductivity $\sigma_{xx}$. In general, the metallic materials fall into the so-called ultraclean regime ($\sigma_{xx} \gtrsim 10^6~(\Omega~cm)^{-1} $). Under such condition, the skew-scattering mechanism should dominate $\sigma_{xy}^{z}$. That is, $\sigma_{xy}^{z} \approx \sigma_{xy}^{z~skew} = S\sigma_{xx}$, in which $S$ is the so-called skewness factor. Therefore, one may expect for the $\sigma_{xy}^{z~extri}$ the relation 
\begin{equation}
	\label{eq:exti}
	\sigma_{xy}^{z~extri} = \sigma_{xy}^{z~skew} + \sigma_{xy}^{z~sj} = S^z \sigma_{xx} + \sigma_{xy}^{z~sj}
\end{equation}

In order to acquire the extrinsic contributions from difference mechanisms, i.e., skew-scattering (skew) and side jump (sj), we then plot $\sigma_{xy}^{z~extri}$ versus the longitudinal conductivity $\sigma_{xx}$ in Fig.~\ref{fig:TaZr} (a). In the dilute limit with $x\leqslant0.1$ or $x\geqslant0.9$, a linear behaviour can be implicitly found. On the basis of Eq.~\ref{eq:exti}, by linearly fitting the data points and extrapolating to $\sigma_{xx}=0$, we get the side-jump contribution $\sigma_{xy}^{z~sj}$ in dilute Ta-Zr alloys. For Ta$_{0.99}$Zr$_{0.01}$ and Ta$_{0.01}$Zr$_{0.99}$, $\sigma_{xy}^{z~sj}$ are approximately 0.27 and -0.11 ($10^{-3}~\Omega~cm$)$^{-1}$, respectively. Compared to the intrinsic contributions (1.63 ($10^{-3}~\Omega~cm$)$^{-1}$ for Ta$_{0.99}$Zr$_{0.01}$ and 1.13 ($10^{-3}~\Omega~cm$)$^{-1}$ for Ta$_{0.01}$Zr$_{0.99}$), the side jump contribution is much smaller. The contribution from skewness scattering mechanism can also be extracted accordingly. We plot in Fig.~\ref{fig:TaZr} (b) the total $\sigma_{xy}^{z}$ with and without vertex corrections (extrinsic contributions), as well as the contribution from skewness scattering as a function of Zr content. It is apparent that in the dilute doping limit, the skewness scattering contributes largely to the total $\sigma_{xy}^{z}$. In contrast, the side jump contribution is independent of concentration and is much smaller than that of skewness scattering. This phenomenon has also been reported by Chadova \textit{et al.} for Pt and Pd hosts doped with 4$d$ or 5$d$ impurities and Cu host with 5$d$ impurities.~\cite{chadova2015tailoring}

In the following, the SHC and SHA dependences on Zr content in Ta$_{1-x}$Zr$_x$ alloys are demonstrated in Fig.~\ref{fig:SHC_SHA} (a) and (b), respectively. For the SHC, we only show $0.05 \leqslant x \leqslant 0.95 $ to make the magnitudes more visible for the concentrated alloys. As can be seen from Fig.~\ref{fig:SHC_SHA} (b) that the maximal $\mid$SHA$\mid$ is around 0.026  with $x=0.4$. At this Zr content, the magnitude of $\mid$SHC$\mid$ is about 1.55 ($10^{-3}~\Omega~\mathrm{cm}$)$^{-1}$, which is higher that of Ta$_{24.9}$Nb$_{18.7}$Hf$_{17.7}$Zr$_{18.3}$Ti$_{20.4}$ (0.29 ($10^{-3}~\Omega~\mathrm{cm}$)$^{-1}$) reported in Ref.~\cite{chen2017spin}.
  For pure bcc Ta, $\sigma_{xy}^{z} = -384~(\Omega~cm)^{-1}$, which is comparable with the value reported by Ref.~\cite{qiao2018calculation} (-284 $(\Omega~cm)^{-1}$, note that due to the different definitions of spin-current operator, the result in Ref.~\cite{qiao2018calculation} has been multiplied by a factor of 2 for the sake of consistency). In this regard, Zr doping in bcc Ta increases $\mid$SHC$\mid$ and $\mid$SHA$\mid$ simultaneously. We have also checked the phase diagram of Ta-Zr and the bcc phase of TaZr alloys can be stabilized.~\cite{okamoto1996ta} Another point to mention is that usually the experimentally measured electric resistivity is higher than the theoretically calculated values. Take bcc Ta as an example, the calculated resistivity is around 0.52 $\mu\Omega~cm$, while the experimental measurements report 50 $\mu\Omega~cm$.~\cite{read1965new,clevenger1992relationship} The possible reason is that the crystal in reality is not perfect, as modelled in the \textit{ab initio} calculations. For alloys, the local distortion is another factor to influence the electric conductivity, which is not considered in the current work. Therefore, we believe that the Ta-Zr binary alloys are rather promising for their prominent spin Hall effects. Finally, one limitation in this work is that only bcc phase is considered. However, using pure Ta as an example, the $\mid$SHC$\mid$ of $\beta$-Ta is around 3 times that of bcc Ta and the $\mid$SHA$\mid$ of $\beta$-Ta is about 10 times that of bcc Ta.~\cite{qiao2018calculation} Nevertheless, this work mainly aims to propose an efficient approach for designing new materials with both high $\mid$SHC$\mid$ and $\mid$SHA$\mid$ facing the large compositional space. Hence, for now we neglect the phase constitution complexity.


\section{Conclusions}

In conclusion, we demonstrate that the MOBO, combined with active learning mediated by density function theory (DFT) calculations, is a highly efficient approach to targeting materials with outstanding SHE in a vast compositional space consisting of Ta, Nb, Hf, Zr and Ti. Using weighted sum UCB as acquisition function, with less than 5 iterations, we find that the TaZr-dominated systems show simultaneously high SHC and SHA. We then focus on the Ta$_{1-x}$Zr$_x$ binary systems and investigate the dependencies of SHC and SHA on the concentration, as well as the intrinsic and extrinsic contributions to the SHC. The maximal $\mid$SHA$\mid$ is around 0.026 in Ta$_{1-x}$Zr$_x$ with $x=0.4$ and the corresponding $\mid$SHC$\mid$ is approximately 1.55 ($10^{-3}~\Omega~\mathrm{cm}$)$^{-1}$, which is comparable to that of Pt ($\sim$4.0 ($10^{-3}~\Omega~\mathrm{cm}$)$^{-1}$).~\cite{kodderitzsch2015linear} The detailed analysis of SHC shows that, especially in the dilute limit, the skewness scattering is the most significant contributor to the large $\mid$SHC$\mid$. The predicted $\mid$SHA$\mid$ (0.026) of Ta$_{0.6}$Zr$_{0.4}$ is also on the same order of magnitude as compare to the measured $\mid$SHA$\mid$ of Pt (0.08).~\cite{liu2011spin}

\begin{acknowledgments}
We appreciate the funding by Deutsche Forschungsgemeinschaft (DFG, German Research Foundation) - Project-ID 405553726 – TRR 270. The Lichtenberg high-performance computer of TU Darmstadt is gratefully acknowledged for providing computational resources for all the calculations carried out in this work. 
\end{acknowledgments}

%

\end{document}